# ROSAT OBSERVATIONS OF COMPACT GROUPS OF GALAXIES

Paolo Saracco and Paolo Ciliegi

Dipartimento di Fisica, Università di Milano,Via Celoria 16, 20133 Milano, Italy

and

Osservatorio Astronomico di Brera,Via Brera 28, 20121 Milano, Italy

saracco@bach.mi.astro.it
ciliegi@bach.mi.astro.it

Abstract

The detection of X-ray emission from Hickson's compact groups of galaxies (HCGs) and its origin are the subject of this work. A search for X-ray emission from compact groups revealed detection from 8 out of the 12 HCG images extracted from the ROSAT public archive. For two of them (HCG16 and HCG44), the X-ray emission originates from point-like sources spatially coincident with galaxies in the group. On the contrary, three groups (HCG62, HCG92 and HCG97) show an extended emission clearly caused by hot intracluster gas. A Raymond-Smith hot plasma model provides an excellent fit to the X-ray spectra. The estimated temperatures are distributed in a quite narrow range (from 0.73 to 0.92 keV) and are consistent, within the errors, with 0.9 keV. The luminosity ranging from 0.75 to $5.1 \cdot 10^{42}$erg s$^{-1}$. The most relevant result is the low metal abundance surely detected in two of them and likely in a third that characterizes the hot gas cloud responsible for the X-ray emission. The data concerning the remaining 3 detected compact groups (HCG4, HCG12 and HCG15) are not sufficient to discriminate with certainty between diffuse and/or point-like X-ray emission. However the results of the spectral analysis point to the presence of a hot gas again with low metal abundance. Our results support the statement that in case of diffuse X-ray emission from compact group of galaxies, the IGM responsible for the emission is predominantly constituted by primordial gas. Upper limits to diffuse emission for the remaining 4 compact groups are given.

1. Introduction

The most common structures in the universe are systems of galaxies like groups and clusters. Thanks to gravity, most galaxies concentrate in groups and clusters (Soneira and Peebles 1978; Ramella et al. 1989) and only a few seem to live in isolation (e.g. Mamon 1993). Groups follow the large scale structure (West 1989; West et al. 1991; Palumbo et al. 1993) and large scale features like the Great Wall (Geller and Huchra 1989; Ramella et al. 1990). In an attempt to obtain a large and complete sample of compact groups of galaxies, Hickson (1982) searched the red prints of the Palomar Observatory Sky Survey (POSS), looking for groups satisfying three well-defined selection criteria. These criteria were chosen to specifically identify compact, isolated systems of galaxies, with the aim of finding true dynamically bounded systems with high galaxy space density. The resulting catalog of compact groups (Hickson 1993; HCGs hereafter) contains systems with apparent space densities even higher than those in the center of rich clusters. Because of their small angular sizes, typically a few arcminutes, it is unlikely that these groups contain many chance projection of field galaxies. On the basis of numerical simulations, Mamon (1990, 1994 and references therein) concluded that chance alignments of galaxies in loose groups could account for most HCGs. From galaxy counts in fields surrounding the groups, Hickson, Kindl and Huchra (1988) estimated that roughly 16% of HCG galaxies could be chance projections, a percentage consistent with field galaxy counts made by Sulentic (1987). However, Palumbo et al. (1994) find that the spiral fraction of HCG galaxies (45%) is significantly lower than that of the surrounding galaxies (54%), a difference that could not be explained in the hypothesis of chance alignments.

Compact groups of galaxies may be part of looser and larger systems (Mamon 1990, 1992). Rood and Struble (1994) detect spatial coincidences between Hickson's compact groups and 36 loose groups and 7 Abell clusters. They conclude that $\sim 70\%$ of HCGs with redshift $z \leq 0.02$ are located within loose groups. Ramella et al. (1994) have recently studied the neighborhoods of 38 HCGs by counting the number of galaxies with measured (accordant) redshifts within 1.5 Mpc. They conclude that 29 of these HCGs are embedded in looser systems although the densities of these neighborhoods are low, typically 8 galaxies within 1.5 Mpc. Studying the spatial distribution of HCG's neighbouring galaxies Palumbo et al. (1994) conclude that more than 80% of the compact groups lie in regions of relatively low densities of galaxies. This has clear implications for the formation of compact groups. Whatever process is responsible for forming compact groups, it must be very efficient in low galaxy density regions. The compact group formation mechanism could act on the majority of galaxies in the immediate surroundings of the group: as a suggestion, large potential wells of dark matter form first. All galaxies in the area of these wells would fall in, producing a compact group before finally merging. The extent of such potential wells would have to be considerably larger than the extent of the present HCGs. X-ray observations are most suitable to investigate the potential well by revealing the extension of the gas surrounding the group. Therefore they are extremely useful in elucidating the nature and dynamical state of these objects and should allow to discriminate between the various compact group formation scenarios (Diaferio et al. 1994).

The first attempts to detect diffuse hot X-ray emitting gas in small groups of galaxies were made by Biermann et al. (1982), Biermann & Kronberg (1983) and Bahcall et al. (1984) with the Einstein Observatory. These observations have detected diffuse hot gas in two compact groups (HCG92 or Stephan's Quintet and Arp330) in the $0.5 - 3$ keV band, but failed to detect X-rays in two other compact groups (HCG40 and HCG34), while in a fifth one the X-ray emission is probably associated with individual galaxies (HCG16; Bahcall et al. 1984). An analysis of the HEAO1 A-2 data by Hickson et al. (1989) produced only upper limits of about $3 \cdot 10^{41}$erg s$^{-1}$ in the $2 - 10$ keV band for 42 HCGs.

Now the ROSAT mission (Trumper 1983) offers a good opportunity to study the X-ray emission from compact groups. Recent high sensitivity ROSAT observations with the Position Sensitive Proportional Counter (PSPC, Pfefferman et al. 1988) have confirmed that some groups of galaxies do have X-ray emission due to the presence of a hot diffuse intergalactic medium. Ebeling et al. (1994) using a complementary source detection algorithm on the short exposures (typically $\sim 10$ min) of the ROSAT All-Sky Survey, detect X-ray emission from 11 Hickson's compact groups. They conclude that the X-ray emission of eight out of 11 groups, point to the presence of a hot inter-group medium (IGM), with luminosity ranging from 0.4 to $8.2 \cdot 10^{42}$erg s$^{-1}$ in the $0.1 - 2.4$ keV energy band. Recently deeper ROSAT PSPC images of some groups were obtained. Mulchaey et al. (1993) detect a hot IGM in the NGC2300 loose group of galaxies. They estimated a hot gas temperature of about 1 keV and an abundance of $\sim 0.06 Z_\odot$ that is the lowest abundance observed for diffuse gas. David et al. (1994) detect an X-ray diffuse emission from the loose group NGC5044, while Ponman & Bertram (1993) detect a large cloud of hot gas centered on the compact group HCG62. The estimated temperature of the cloud is about 1 keV and the metal abundance $Z \sim 0.15 Z_\odot$. Such low metallicity suggests that the intragroup medium is predominantly primordial gas with some galactic gas intermixed. On the other hand Sulentic et al. (1994) obtain different results on HCG92 (Stephan's Quintet): they detect a hot diffuse X-ray emission from a gas cloud that seem to have solar abundance. Therefore an extended and detailed study of X-ray emission from HCGs is fundamental to know the dynamical properties of compact systems of galaxies and their formation history as well as their own way to evolve.

Due to the large ROSAT/PSPC field of view (∼50 arcmin radius) some HCGs were serendipitously observed by ROSAT. In order to study X-ray properties of HCGs in a systematic way, we searched in the ROSAT public archives for all the ROSAT/PSPC observations containing a Hickson's compact group within the field of view. In this paper we report the results of the analysis of the data thus collected. In section 2, we describe the sample and the selection criteria, in section 3 the data reduction and analysis procedures; in section 4, the results and, finally, section 5 contains discussion and conclusions. Throughout this paper, we adopt $H_0 = 50$ km s$^{-1}$Mpc$^{-1}$.

## 2. The Sample

Using the ROSAT public archives available on 1994 March 31, we searched all the ROSAT/PSPC observations containing a Hickson's compact group within the field of view. To be included in our sample, the observation must have an exposure time greater than 3000 seconds and an off-axis angle $\theta$ (the distance between the center of the ROSAT/PSPC field of view and the position of the source) smaller than 48 arcmin.

We have found 13 ROSAT/PSPC observations obeying such criteria. Seven compact groups (HCG 4,12, 16, 44, 62, 92 and 97) were pointed targets while the other 6 compact groups (HCG 3, 15, 35, 38, 57 and 73) fell serendipitously in the field of view. In Table 1, we list the basic data of these 13 Hickson's compact group of galaxies: in column 1 the source name is given, in column 2 the redshift z, in columns 3 and 4 the radial velocity dispersion and the mass-to-light ratio respectively (Hickson et al. 1992), in column 5 the Galactic neutral hydrogen column density along the line of site ($cm^{-2}$) from Stark et al. (1992), the identification number (ROR) and the observing time (Obs. Time) of the ROSAT/PSPC observation in columns 6 and 7 and, finally, in column 8 the off-axis angle $\theta$ (arcmin). Unfortunately, HCG35 is covered by the window support structure of the PSPC instrument, so it was excluded by the sample. Thus our final sample contains 12 Hickson's compact groups. It must be remarked that while the sample size is of the same order than that of Ebeling et al. (1994), the observations we used are much deeper than the ROSAT All-Sky Survey exposures.

### 3. Data Analysis

All the PSPC observations were performed with the PSPC-B in the focal plane of the ROSAT telescope. We fitted each source with a two-dimensional gaussian function to determine its centroid. Subsequently we extracted the source counts in the 0.5-2.3 keV band from a circular area centered on the source centroid. The extraction radius $R$ is given by the distance from the centroid where the radial background subtracted profile of the source meets the background level. The background counts and dispersion were estimated in an annulus centered on the source. When detected, other sources within the source or background region were removed using a 1.5′ radius circle.

For each group, a contour level map has been produced in the 0.9-1.1 keV energy range. The contours were generated by smoothing the images with a gaussian filter of about 30 arcsec FWHM. This is the approximate resolution of the X-ray telescope/PSPC combination in the center of the field of view in case of photons of about 1 keV. The lowest contour corresponds to $1\sigma$ above the background. The other contours have been chosen to optimize the plots. The heavy line circle showed in each figure represents the point spread function (PSF) of the ROSAT telescope at 1 keV and at the corresponding off-axis angle (Hasinger et al. 1994), while the crosses represent the optical positions of the single galaxies marked as in Hickson (1993).

The data were processed using the Standard Analysis Software System (SASS). The 34 energy channels from the MPE SASS pipeline were used for the spectral analysis. The first twelve channels ($<0.5$ keV) were excluded from the final spectra because the low-energy spectrum is substantially affected by uncertainties in background subtraction (Snowden et al. 1994) while the last channel ($>2.4$ keV) was excluded because the response of the instrument is not well defined at the extremes of the energy range. Finally the spectrum was compressed in order to obtain unequally spaced bins each containing sufficient counts and a signal to noise ratio S/N$>3$, so that the $\chi^2$ statistic can be applied. Model fits were carried out using the XSPEC software package (version 8.34) and the best-fit model parameters were obtained by $\chi^2$ minimization. Following Fiore et al. (1994), we have used the 1992 March response matrix for observations made before October 1991 and the 1993 January response matrix for observations made after October 1991.

In case of presumed diffuse emission we compared the results obtained by fitting the data with a Raymond and Smith (1977; RS hereafter) hot plasma model at solar abundance ($Z = 1Z_\odot$, indicated as fit 1 in Table 3) with those obtained considering $Z$ as a free parameter (indicated as fit 2). Besides the $\chi^2$ test on goodness of fit, we measure the statistical significance of the reduction in $\chi^2$ with the addition of $Z$ as a free parameter using the F-test of additional term (Bevington and Robinson 1992). For each compact group we have constructed the statistic
$$F_{\chi^2} = \frac{\chi^2(n) - \chi^2(n+1)}{\chi^2(n+1)/(N-n-1)}$$
($n$ is the number of parameters) that follows the $F$ distribution with $\nu_1 = 1$ and $\nu_2 = N - n - 1$, and we calculated the probability $P(F > F_{\chi^2})$ that the reduction in $\chi^2$ is not statistically significant. In case of sufficient counts we also test the improvement of the $\chi^2$ obtained by adding $N_H$ as a free parameter (fit 3).

## 4. Results

Out of the 12 selected HCGs, 8 have been detected at $> 5\sigma$ level. The X-ray emission from two of them (HCG 16 and 44) is clearly due to point-like sources spatially coincident with group galaxies. Of the remaining 6 compact groups, some of them show a clearly X-ray diffuse emission (HCG 62, 92 and 97) while in others the presence of a hot gas cloud is not so obvious (HCG 4, 12 and 15). In the following we will report in turn the results of the analysis obtained for each group. Table 2 summarizes the results concerning X-ray emission from the single galaxies of HCG 16 and 44. Table 3, in which we report the results of the spectral analysis of the remaining detected HCGs, is organized as follows: group catalog number (column 1); extraction radius R (column 2); net counts and $1\sigma$ error (column 3); code number indicating the three different fits (see §3; column 4); neutral hydrogen column density, metallicity and temperature with relative $1\sigma$ error (columns 5, 6 and 7); $\chi^2$ value and associated probability (columns 8 and 9); the probability $P(F > F_\chi^2)$ that the fit $n$ does not provide a significantly better fit with respect to fit $n-1$ (column 10), flux and luminosity in the 0.5-2.3 keV band (columns 11 and 12). Finally in Table 4 we report the flux and luminosity $5\sigma$ upper limit of the diffuse emission for the undetected sources.

### 4.1 Detected Compact Groups

*HCG4*

This group consists of triplet with accordant redshift (galaxies A, C and D) plus two galaxies with discordant redshifts that are not included in fig. 1. The brightest galaxy (spiral) is an infrared source. In the 9328 s exposure the PSPC on ROSAT detected $\sim 1470$ counts from a source centered on the galaxy group. Fig. 1 shows the contour map. The map shows an X-ray emission apparently centered at a location corresponding to the position of the brightest galaxy and an extension comparable to the PSF. A possible interpretation is that the X-ray emission comes from the dominant galaxy: integrated emission from individual sources inside the galaxy itself or from hot gas associated with the interstellar medium of the galaxy (Fabbiano et al 1992). On the other hand we cannot exclude the presence of an intragroup medium: the exposure time was not sufficient to determine whether the IGM of this group is just too cool and/or diffuse to be seen, or is less abundant. With fit 1 (assuming solar metallicity) we do not obtain a good fit ($\chi^2/dof = 83/16$), while a resonable result ($\chi^2/dof = 14/15$) is obtained with fit 2 ($kT = 0.97 \pm 0.1$ keV, $Z < 0.02 Z_\odot$). In fig. 2, 68%, 90% and 99% confidence level contours for the RS model fit are shown. This result seems to point to the presence of a hot IGM rather than emission from a single galaxy. The addition of $N_H$ as a free parameter (fit 3) does not significantly improve the fit. The estimated value of $N_H$ is consistent with the Galactic hydrogen column density. We adopt the low value of the metal abundance in estimating the X-ray energy flux and luminosity in the 0.5-2.3 keV band. We estimate a flux $F_X = 1.86 \cdot 10^{-12}$erg cm$^{-2}$s$^{-1}$ and a luminosity $L_X \sim 6.3 \cdot 10^{42}$ erg s$^{-1}$.

*HCG12*

This group consists of a quintet of mostly S0 type galaxies with accordant redshift. It has a relatively low value of mass to light ratio. In the 10789 s exposure, the PSPC detected only $84 \pm 15$ counts in the 0.5-2.3 keV band. Fig. 3 shows the contour plot. Also, in this case, the extension of the emission is comparable with the PSF. Moreover, the low counts detected are not sufficient to fit the data and to try to discriminate between diffuse emission or emission from single galaxies, although the X-ray map shows an emission centered at a location corresponding to the brightest galaxy (A) of the group. We binned the data in a single channel and we estimated flux and luminosity assuming $N_H = N_{Hgal}$ for metallicities $Z = 1 Z_\odot$ and $Z = 0.1 Z_\odot$ and temperatures $kT = 1.0$ keV and $kT = 0.2$ keV. For all choices of temperature and metallicity, the flux remaines roughly constant at a value of $\sim 1.0 \cdot 10^{-13}$erg cm$^{-2}$s$^{-1}$ (see Table 3).

*HCG15*

This is a sextet of early and late type galaxies with a high radial velocity dispersion that contributes to the unusually high mass-to-light ratio. Galaxies A and D are radio sources and D is interacting with F. This group was not observed as a target and its off-axis angle is about 34' (Table 1). Within a radius $R = 6'$ ROSAT detected $\sim 400$ counts in the 0.5-2.3 keV energy band in 14000 s exposure. The contour map (fig. 4) shows an irregularly shaped source. The emission seems to encompass the entire sextet although clumps are present corresponding to the two interacting galaxies (D and F) and to the A and E galaxies. The RS model with a constrained metallicity value at solar abundance (fit 1) gives a $\chi^2/dof = 10.86/9$ while we obtained a $\chi^2/dof = 3.98/8$ with unconstrained metallicity (fit 2). The significance of the improvement of the reduced chi-squared value is proved by the low value of the probability $P(F > F_\chi) = 0.005$. The estimated metal abundance is $Z < 0.19 Z_\odot$ and the temperature of the emitting gas is $kT = 1 \pm 0.38$ keV. We estimate a flux $F_X \sim 4.5 \cdot 10^{-13}$ erg cm$^{-2}$ s$^{-1}$ and a luminosity $L_X = 10^{42}$ erg s$^{-1}$.

*HCG16*

This group consists of four late type galaxies of comparable luminosity. All but galaxy B have strong radio and infrared emission (Menon & Hickson 1985; Menon 1992). It has a low radial velocity dispersion and correspondingly a very low mass-to-light ratio. The contour map (fig. 5) shows emission clearly due to point like sources. The image shows at least three distinct peaks corresponding to the position of galaxies A+B, C and D. The net counts detected are 184 (galaxies A and B), 196 (C) and 80 (D) respectively (see Table 2). We estimated the flux and the luminosity assuming solar abundance ($Z = 1 Z_\odot$) and a temperature $kT = 1$ keV. An upper limit to a diffuse emission due to a possible presence of a hot gas cloud is estimated too (see Table 4). In this case, we assume a metallicity $Z = 0.1Z_\odot$ and a temperature $KT = 1$ keV. Fig. 5 shows other two X-ray sources (XR1 and XR2) that are not apparently related to the group. In order to identify these sources, we searched for possible counterparts in the Radio Master Catalog and in the Guide Star Catalog. The search was performed by using the HEASARC Online Service that is provided by NASA-Goddard Space Flight Center. Moreover we searched for possible extragalactic counterparts in the NASA/IPAC Extragalactic Database. We have not found any known source within 30″ from the position of XR1 and XR2. Therefore, only an optical spectroscopic observation can be useful to identify these sources and can show whether they are related or not to the group. In Table 5, we report the positions and count rates of these two unclassified sources.

*HCG44*

This is a nearby group of three spiral galaxies (A, C and D) with peculiar morphology and radio emission (Menon & Hickson 1985; Menon 1992), plus an elliptical galaxy (B). The radial velocity dispersion and the mass to light ratio are not high. As for HCG16, the X-ray emission is due to point-like sources. From fig. 6, two peaks of emission can be seen at locations corresponding to position of the galaxies A and B with net counts 29 and 35 respectively. Galaxies C and D have not been detected. Under the same assumption of metallicity and temperature made for HCG16, flux and luminosity are estimated for galaxies A and B as well as upper limits to the emission of galaxies C and D (Tab. 2). In table 4, we report the upper limit to the diffuse emission of HCG44 with an assumed metal abundance $Z = 0.1Z_\odot$ and $kT = 1$ keV. Also in this case there are 4 X-ray sources (XR1, XR2, XR3 and XR4) that seem to be unrelated to the group (fig. 6).

Using the same procedure adopted to classify the sources in the field of HCG16, we have not found any known counterparts of these four sources. In Table 5, right ascension, declination and count rate of the unclassified sources are listed.

*HCG62*

This group consists of four early type galaxies one of which is a compact nuclear radio source. The velocity dispersion and mass to light ratio are relatively large. It was observed as a target and it has been studied quite carefully by Ponman & Bertram (1993). In the 19715 s exposure the PSPC on ROSAT detected $\sim 11000$ counts from a source centered on the galaxy group. The contour plot in the 0.9-1.1 keV band (fig. 7) shows an extended emission encompassing all the members of the group. The fixed metallicity fit of the integrated source spectrum within $R = 15'$ does not give a satisfactory result whilst, leaving $Z$ as a free parameter, we obtain quite a good fit ($\chi^2/dof = 14.6/18$) with an estimated temperature of $0.92 \pm 0.02$ keV and an abundance $Z = 0.15 \pm 0.07 Z_\odot$. Fig. 8 shows the confidence level contours for the RS model fit. The estimated flux and luminosity in the 0.5-2.3 keV energy band are $F_X = 6.1 \cdot 10^{-12}$ erg cm$^{-2}$s$^{-1}$ and $L_X = 4.96 \cdot 10^{42}$ erg s$^{-1}$. The addition of $N_H$ as a free parameter (fit 3) marginally improves the fit. However it must be remarked that the estimated $N_H$ is consistent with $N_{Hgal}$. The temperature, obtained from RS model fits to spectra from successive rings about the X-ray centroid, dips sharply in the center ($R < 4'$) and decreases steadily with increasing radius (fig. 9). These results are in quite good agreement with those obtained by Ponman & Bertram (1993). The contrast between the low metallicity value obtained and that expected in gas ejected from galaxies ($Z \geq 1 Z_\odot$), points to the presence of a hot intragroup medium predominantly constituted by primordial gas.

### HCG92

This is the well known Stephan's Quintet (SQ), also known as Arp 319. It consists of a bright foreground spiral galaxy (A) plus four late type galaxies. One of these is an infrared and radio source. This group has been detected at first by the IPC on the Einstein Observatory and studied by Bahcall et al. (1984) and recently by Sulentic et al. (1994) on the basis of a Rosat/PSPC observation.

This 20869 s observation recorded $\sim$ 770 photons from the group. The contour map (fig.10) shows an irregularly shaped source and an extended emission encompassing all the group's members. The RS model with solar gives $\chi^2/dof = 6.42/10$ while we obtain $\chi^2/dof = 4.62/9$ with unconstrained metallicity, yielding a low metal abundance ($Z = 0.2 \pm 0.2$). In this case, the improvement of the reduced chi-square value is only marginally significant. In fig. 11, the confidence level contours of the RS model fit are given. The estimated flux and luminosity are $F_X = 3.8 \cdot 10^{-13}$ erg cm$^{-2}$ s$^{-1}$ and $L_X = 7.5 \cdot 10^{41}$ erg s$^{-1}$. Also, in this case, the addition of $N_H$ as a free parameter does not significantly improve the fit, yielding a value of $N_H$ consistent, within the errors, with $N_{Hgal}$.

### HCG97

This group is a loose quintet with early and late type galaxies and it was observed by Rosat as a target. In the 13886 s exposure, the PSPC on ROSAT detected $\sim$ 2960 counts from a source centered on group. The contour plot (fig. 12) clearly shows an extended emission and an irregularly shaped source. The RS model fit to the integrated source spectrum within 15′ with solar metallicity is bad.

On the contrary, a satisfactory result is obtained leaving $Z$ as a free parameter. In this case, the fit gives a temperature $kT = 0.72 \pm 0.1$ keV and metallicity $Z < 0.02 Z_\odot$. In fig. 13 the confidence level contours for the fit are given. The estimated flux and luminosity in the 0.5-2.3 keV band are $F_X = 2.5 \cdot 10^{-12}$ erg cm$^{-2}$ s$^{-1}$ and $L_X = 5 \cdot 10^{42}$ erg s$^{-1}$. The fit obtained by adding $N_H$ as a free parameter is not significantly better than fit 2. The estimated $N_H$ is consistent with $N_{Hgal}$. Given the high number of net counts, we have extracted counts from annuli of 3 arcmin width about the centroid and fitted the data with a RS model. The best fit temperature drops steadily with increasing radius as show in figure 14. Also, the metallicity, although it is less well defined, seems to decrease with increasing radius. However, it is clear that the metal abundance is modest and therefore that the IGM must be predominantly primordial gas.

In order to test how much the exclusion of the ROSAT data in the energy range 0.1 to 0.5 keV could affect the results, we have repeated the spectral analysis of HCGs 4, 62 and 97 by considering this energy range. The estimated temperatures are not more than 20% lower than those obtained by fitting the data in the 0.5–2.4 keV energy range and the metallicity is always very low. However, in all cases we do not obtain a satisfactory fit showing that the data are not reliable in this energy range (Snowden et al. 1994).

### 4.2 Undetected Compact Groups

For the remaining four non detected compact groups of galaxies (HCGs 3, 38, 57 and 73) that were not observed as targets and that were not even detected by Ebeling et al. (1994), we estimated upper limits to the diffuse X-ray emission. To calculate the upper limits, on the basis of previous results, we assumed a low metallicity value ($Z = 0.1 Z_\odot$) and a temperature $kT = 1$ keV. All but one compact group (HCG73) should have an X-ray luminosity less than $10^{42}$ erg s$^{-1}$ in the 0.5-2.3 keV energy band. The results are summarized in Table 4 in which we report: HCG number (column 1); flux (column 2) and luminosity (column 3).

## 5. Discussion and Conclusions

We have revealed X-ray emission from 8 out of the 12 Hickson's compact groups of galaxies searched in the ROSAT public archives available at 1994. Two of them (HCG16 and HCG44), that were missed by Ebeling et al. (1994) in their analysis of the All-Sky Survey, clearly show an emission due to point-like sources coincident with the positions of members of groups. HCG16 had been detected by Bahcall et al. (1984) with the Einstein Observatory. From these early data, the source, with an estimated luminosity of about $2 \cdot 10^{41}$ erg s$^{-1}$, seemed to be centered (within the IPC uncertainty) on the central galaxy NGC383 and to be extended in the direction of the other galaxies. Moreover the velocity dispersion of the group is too small in comparison with the high temperature implied by the X-ray emission. On the basis of these considerations, Bahcall et al. concluded that most of the X-ray emission from HCG16 originates in the galaxies themselves. With the ROSAT PSPC, it has been possible not only to check this result, but also to detect the single sources within HCG16. At least three out of four galaxies (A or B, C and D) are X-ray sources. It has not been possible to resolve galaxies A and B because of their small separation. Galaxy C (NGC383) is the dominant X-ray component with a luminosity of $1.5 \cdot 10^{41}$ erg s$^{-1}$. In HCG44, only two galaxies clearly show X-ray emission: the spiral galaxy A and the elliptical galaxy B. Both have an X-ray luminosity at least one order of magnitude lower than galaxies in HCG16. The other two galaxies (C and D), which are fainter than A and B in the visual band (Hickson 1993), have not been detected by ROSAT and their X-ray luminosity should be less than $2 \cdot 10^{39}$ erg s$^{-1}$ (see Table 2).

Three compact groups (HCG 62, 92 and 97) clearly show X-ray diffuse emission. For HCG62 we obtain results in very good agreement with those obtained by Ponman & Bertram. The low metallicity value obtained by fitting the data with a RS model confirms the presence of a hot gas cloud predominantly constituted by primordial gas. In HCG97, the estimated temperature of the gas cloud is lower than that of HCG62. While in the latter the temperature dips sharply in the centre suggesting the presence of a central cooling flow (Ponman & Bertram 1993), in the former it tends to decrease steadily with increasing radius. Therefore, we exclude, on the basis of the temperature trend, the presence of a central cooling flow in HCG97 while we confirm the results of Ponman and Bertram for HCG62. Also in HCG97 the estimated metal abundance points to a low value showing the presence of an IGM predominantly formed by primordial gas. An alternative explanation for the low metallicity is that most of gas ejected from the galaxies may have escaped from their group probably at early times (Renzini 1994).

In HCG92 (SQ), the X-ray emission clearly originates mainly in a hot gas cloud encompassing all the group galaxies, while the abundance of the IGM is less well defined. The first X-ray detection of this group is that of Bahcall et al. (1984) who concluded that the X-ray emission originates in a hot IGM ($kT \sim 0.4$ keV). This conclusion was supported by the fact that the estimated luminosity of about $10^{42}$ erg s$^{-1}$ in the 0.5-3 keV band is higher than that usual for galactic emission. In addition, the emission is soft (∼ 80% of counts fall in the low energy channels ≤ 1 keV), while X-ray spectra of late-type galaxies generally show a much smaller fraction of their emission in these low-energy bands (Fabbiano et al. 1982). Finally, the emission appears extended and the centroid is consistent with the group center. Very recently, Sulentic et al. (1994) have observed HCG92 with ROSAT detecting diffuse X-ray emission from a source with a diameter of about 6.5′ centered on the group. They argue that the emitting gas cloud is likely to have a near solar abundance rather than a primordial value. They compare the results of the spectral analysis obtained by fitting power-law, thermal bremsstrahlung and RS models to the observed spectrum. They obtain a resonable result only with an RS model ($\chi^2/dof = 1.2$) when the value of $N_H$ is constrained near galactic values. The metallicity is never considered as a free parameter and they always assume a nearly solar abundance. On the basis of the residuals of the bremsstrahlung model fit, that show a deficiency in the energy region where the largest concentration of emission lines are found, and the much better fit obtained with the RS model at solar abundance, they conclude that the IGM has a nearly solar metallicity rather than a primordial value. In our analysis, we compare the results obtained by fitting RS model to the observed spectrum in the cases where the metallicity is solar and when it is a free parameter. In both cases, we obtain resonable results as the $\chi^2$ statistics shows. Although in this case, the F-test gives an only marginally significant result, the low probability is suggestive of a real difference between the two $\chi^2$ values and the presence of an IGM mainly constituted by low metallicity gas cannot be excluded.

However, it must be remarked that the X-ray properties of HCG92 show some differences with respect to those of HCG62 and HCG97. In these two groups, the X-ray emission extends far ($R \geq 15'$) beyond the optical radius of the groups ($R^{opt}_{HCG62} \sim 1.9'$ and $R^{opt}_{HCG97} \sim 2.6'$; Hickson 1982) while in HCG92 the extension of the X-ray emitting cloud ($R \sim 3'$) is comparable with its optical angular diameter ($D \sim 3.2'$). Moreover, while the X-ray luminosities of HCG62 and HCG97 are comparable and they are of about $5 \cdot 10^{42}$ erg s$^{-1}$, the luminosity of HCG92 is one order of magnitude lower. We cannot exclude the hypothesis of a different origin between the IGM in HCG92 and that in HCG62 and HCG97. Evidence for dynamical interaction between the galaxies in group could support this statement. Two large clouds of neutral hydrogen at the velocity of the group have been detected by Allen & Sullivan (1980) and they could represent gas stripped from the group galaxies.

The remaining 3 detected compact groups of galaxies (HCG4, HCG12 and HCG15), two of which (12 and 15) have been missed by Ebeling et al. (1994), do not show sufficient X-ray emission to make a reliable interpretation. In all cases, the extension of the emission is comparable to the PSF. Although it has been detected at $> 5\sigma$, HCG12 does not have sufficient counts to fit the spectra. It must be remarked that the exposure times for these three groups is on average, half of those of HCG62 and HCG92, while their redshifts are always larger, so that the exposure time was not sufficient to investigate whether or not a hot gas cloud is present in these compact groups. Indeed, the lower detectable surface brightness (counts arcmin$^{-2}$s$^{-1}$) in the exposures of HCGs 4, 12 and 15, are at least one half of those of HCGs 62 and 97. However, the estimated X-ray luminosities ($\geq 10^{42}$h$_{50}^{-2}$erg s$^{-1}$) are higher than is usual for galactic emission (Fabbiano et al.

1992). The X-ray emission from HCG4 seems to be regular and centered on the brightest galaxy A (fig. 1), and its luminosity may be understood if galaxy A were an AGN as suggested by Ebeling et al. However we need optical spectroscopic data in order to unambiguously discriminate on the nature of this galaxy. On the contrary, the contour level map of HCG15 (fig. 4), shows clumps in correspondence with galaxies A and E, and in correspondence with the two interacting galaxies (D and F). Its luminosity, cannot be explained by emission from individual galaxies. Moreover, the RS hot plasma model fits to the spectra of HCG4 and HCG15 give reasonable results for low metallicity (see table 3). It must be remarked that in all cases in which we considered $N_H$ as a free parameter (HCGs 4, 62, 92 and 97) there is not a significant improvement of fit. The estimated metallicity and temperature are unchanged with respect to fit 2. Moreover, the estimated $N_H$ is always consistent with the Galactic value, suggesting that the intrinsic absorption is negligible in these sources. Therefore, our data suggest the presence of a diffuse emission originating in a relatively hot IGM characterized by low metallicity.

## AKNOWLEDGEMENTS

Our thanks go to G. Chincarini, B. Garilli, A. Iovino, T. Maccacaro, D. Maccagni and A. Wolter for offering helpful discussions and suggestions which improved the manuscript. We would like to thank S. Campana for offering discussions while this work was in progress. The referee G. Mamon made suggestions which led to substantial improvements in the presentation of these results.

West M.J., Villumsen J. V. & Dekel A. 1991, ApJ 369, 287

FIGURE CAPTIONS

Figure 1. X-ray contour map at 1 keV of HCG4. The image was smoothed with a gaussian function with FWHM=30″. The heavy line circle represents the 1 keV Rosat point spread function (PSF) at the same off-axis angle of the source. The crosses represent the optical position of the single galaxies marked as in Hickson (1993). The contour levels are $1\sigma$, $2\sigma$, $4\sigma$, $8\sigma$, $70\sigma$, $140\sigma$, $180\sigma$ above the background.

Figure 2. 68%, 90% and 99% confidence levels for the RS model fit to spectra of HCG4.

Figure 3. Same as in Fig. 1 for HCG12. The contour levels are $1\sigma$, $2\sigma$, $3\sigma$, $4\sigma$ and $5\sigma$ above the background.

Figure 4. Same as in Fig. 1 for HCG15. The contour levels are $1\sigma$, $1.5\sigma$, $2\sigma$, $2.5\sigma$, $3\sigma$, $4\sigma$, $5\sigma$ and $6\sigma$ above the background.

Figure 5. Same as in Fig. 1 for HCG16. The contour levels are $1\sigma$, $1.5\sigma$, $2.5\sigma$, $4\sigma$, $7\sigma$, $11\sigma$ and $17\sigma$ above the background.

Figure 6. Same as in Fig. 1 for HCG44. The contour levels are $1\sigma$, $1.5\sigma$, $2\sigma$, $2.5\sigma$, $3\sigma$, $3.5\sigma$, $4\sigma$, $5\sigma$, $5.5\sigma$ and $6\sigma$ above the background.

Figure 7. Same as in Fig. 1 for HCG62. The contour levels are $1\sigma$, $1.5\sigma$, $2\sigma$, $2.5\sigma$, $3\sigma$, $6\sigma$, $12\sigma$, $21\sigma$, $35\sigma$, $50\sigma$ and $62\sigma$ above the background.

Figure 8. Same as in Fig. 2 for HCG62

Figure 9. Temperature of HCG62 from RS model fits to spectra from annuli of 2.5′ width about the X-ray centroid. Errors are 1$\sigma$.

Figure 10. Same as in Fig. 1 for HCG92. The contour levels are 1$\sigma$, 1.5$\sigma$, 2.5$\sigma$, 5$\sigma$, 8$\sigma$, 13$\sigma$, 16$\sigma$, 19$\sigma$, 21$\sigma$ and 24$\sigma$ above the background.

Figure 11. Same as in Fig. 2 for HCG92

Figure 12. Same as in Fig. 1 for HCG97. The contour levels are 1$\sigma$, 1.5$\sigma$, 2.5$\sigma$, 5$\sigma$, 8$\sigma$, 11$\sigma$, 13$\sigma$, 17$\sigma$, 20$\sigma$, 23$\sigma$ and 28$\sigma$ above the background.

Figure 13. Same as in Fig. 2 for HCG97

Figure 14. Temperature of HCG97 from RS model fits to spectra from annuli of 3′ width about the X-ray centroid. Errors are 1$\sigma$.

Table 1: Basic data of the 13 selected HCG observations extracted from the ROSAT public archive.

| Source | $z$ | $\sigma_v$ (km s$^{-1}$) | M/L (M$_\odot$/L$_\odot$) | $N_{H_{gal}}$ ($10^{20}$cm$^{-2}$) | ROR | time$_{exp}$ (s) | $\theta$ (arcmin) |
|---|---|---|---|---|---|---|---|
| HCG 3 | 0.0255 | 251 | 182 | 3.49 | WP400178 | 7905 | 13.6 |
| HCG 4 | 0.0280 | 339 | 115 | 1.55 | WP800356 | 9328 | 0.3 |
| HCG 12 | 0.0485 | 240 | 37 | 4.5 | WP800358 | 10789 | 0.4 |
| HCG 15 | 0.0228 | 427 | 309 | 2.99 | RP700432 | 14070 | 33.8 |
| HCG 16 | 0.0132 | 123 | 11 | 2.18 | RP800114 | 14853 | 0.8 |
| HCG 35 | 0.0542 | 316 | 41 | 2.79 | RP900009 | 15703 | 41.9 |
| HCG 38 | 0.0292 | 13 | —- | 3.43 | WP700471 | 3892 | 35.8 |
| HCG 44 | 0.0046 | 135 | 40 | 2.13 | RP800125 | 4671 | 0.6 |
| HCG 57 | 0.0304 | 269 | 35 | 1.86 | WP700476 | 4819 | 30.3 |
| HCG 62 | 0.0137 | 288 | 68 | 3.00 | WP800098 | 19715 | 0.6 |
| HCG 73 | 0.0449 | 123 | 47 | 3.48 | RP700312 | 4557 | 48.2 |
| HCG 92 | 0.0215 | 389 | 22 | 7.70 | WP800066 | 20879 | 4.0 |
| HCG 97 | 0.0218 | 372 | 174 | 3.62 | WP800357 | 13886 | 0.7 |

**Table 2:** X-ray properties of emission from single galaxies. Flux and luminosity have been estimated assuming solar abundance ($Z = Z_\odot$) and a temperature $kT = 1$ keV. Upper limits to the X-ray emission of undetected galaxies C and D of HCG44 are given.

| Galaxy | Net Counts | $F_{X(0.5-2.3\ \text{keV})}$ $(10^{-13}\text{erg cm}^{-2}\text{s}^{-1})$ | $L_{X(0.5-2.3\ \text{keV})}$ $(10^{40}\text{erg s}^{-1})$ |
|---|---|---|---|
| HCG 16 | | | |
| A+B | 184 ± 19 | 1.31 ± 0.1 | 9.78 |
| C | 196 ± 20 | 1.40 ± 0.1 | 10.5 |
| D | 80 ± 16 | 0.57 ± 0.01 | 4.26 |
| HCG 44 | | | |
| A | 29 ± 12 | 0.52 ± 0.02 | 0.49 |
| B | 35 ± 13 | 0.52 ± 0.02 | 0.49 |
| C | ——— | < 0.24 | < 0.23 |
| D | ——— | < 0.24 | < 0.23 |

**Table 3:** Results of spectral analysis of X-ray emission from HCGs.

| Source | $R$ (arcmin) | Net Counts (0.5-2.3 keV) | Fit | $N_H$ ($10^{20}$cm$^{-2}$) | $Z$ $Z_\odot$ | $kT$ (keV) | $\chi^2/dof$ | $P(\chi^2)$ | $P(F>F_{\chi^2})$ | $F_{X(0.5-2.3\ \mathrm{keV})}$ ($10^{-12}$erg cm$^{-2}$ s$^{-1}$) | $L_{X(0.5-2.3\ \mathrm{keV})}$ ($10^{42}$erg s$^{-1}$) |
|---|---|---|---|---|---|---|---|---|---|---|---|
| HCG4 | 2.5 | $1470\pm44$ | 1 | 1.55 | 1.00 | $2.83^{+0.97}_{-0.68}$ | 82.74/16 | 0.00 | | $2.03\pm0.06$ | 6.80 |
| | | | 2 | 1.55 | $<0.02$ | $0.97^{+0.10}_{-0.10}$ | 14.00/15 | 0.53 | $2\cdot10^{-7}$ | $1.86\pm0.06$ | 6.30 |
| | | | 3 | $1.31\pm0.50$ | $<0.07$ | $0.97\pm0.34$ | 13.67/14 | 0.47 | $59\cdot10^{-2}$ | $1.81\pm0.06$ | 6.10 |
| HCG12* | 2.5 | $84\pm15$ | | 4.50 | 1.00 | 1.00 | | | | $0.09\pm0.02$ | 0.89 |
| | | | | 4.50 | 0.10 | 1.00 | | | | $0.10\pm0.02$ | 1.01 |
| | | | | 4.50 | 1.00 | 0.20 | | | | $0.14\pm0.03$ | 1.40 |
| | | | | 4.50 | 0.10 | 0.20 | | | | $0.13\pm0.02$ | 1.31 |
| HCG15 | 6.0 | $398\pm36$ | 1 | 2.99 | 1.00 | $>1.62$ | 10.86/9 | 0.29 | | $0.53\pm0.05$ | 1.19 |
| | | | 2 | 2.99 | $<0.19$ | $1.00^{+0.38}_{-0.38}$ | 3.98/8 | 0.86 | $5\cdot10^{-3}$ | $0.45\pm0.04$ | 1.01 |
| HCG62 | 15 | $11097\pm237$ | 1 | 3.00 | 1.00 | $1.00^{+0.02}_{-0.02}$ | 64.93/19 | 0.00 | | $5.73\pm0.02$ | 4.72 |
| | | | 2 | 3.00 | $0.15\pm0.07$ | $0.92^{+0.02}_{-0.02}$ | 14.6/18 | 0.69 | $5\cdot10^{-7}$ | $6.14\pm0.01$ | 5.06 |
| | | | 3 | $3.32\pm1.00$ | $0.09^{+0.12}_{-0.06}$ | $0.86^{+0.14}_{-0.18}$ | 11.7/17 | 0.82 | $5.6\cdot10^{-2}$ | $7.04\pm0.15$ | 5.80 |
| HCG92 | 3 | $769\pm39$ | 1 | 7.70 | 1.00 | $0.82^{+0.05}_{-0.08}$ | 6.42/10 | 0.78 | | $0.43\pm0.02$ | 0.86 |
| | | | 2 | 7.70 | $0.21^{+0.20}_{-0.20}$ | $0.76^{+0.10}_{-0.15}$ | 4.62/9 | 0.87 | $9\cdot10^{-2}$ | $0.38\pm0.02$ | 0.76 |
| | | | 3 | $10.37\pm2.70$ | $0.21\pm0.32$ | $0.73\pm0.20$ | 4.60/8 | 0.80 | $86\cdot10^{-2}$ | $0.52\pm0.03$ | 1.03 |
| HCG97 | 15 | $2968\pm76$ | 1 | 3.62 | 1.00 | $0.97^{+0.03}_{-0.03}$ | 140.7/16 | 0.00 | | $2.15\pm0.06$ | 4.41 |
| | | | 2 | 3.62 | $<0.02$ | $0.77^{+0.08}_{-0.09}$ | 17.11/15 | 0.31 | $10^{-7}$ | $2.49\pm0.06$ | 5.10 |
| | | | 3 | $1.00^{+7.40}_{-1.00}$ | $<0.04$ | $0.85\pm0.16$ | 16.38/14 | 0.29 | $44\cdot10^{-2}$ | $2.63\pm0.10$ | 5.40 |

*Given the low net counts, we binned the data in a single channel and we estimated flux (assuming $N_H=N_{H_{gal}}$) for $Z=1.0Z_\odot$ and $Z=0.1Z_\odot$, fixing $kT$ to 0.2 and 1.0 keV.

**Table 4:** Upper limits to diffuse X-ray emission from undetected groups. Flux and luminosity have been estimated assuming $Z = 0.1 Z_\odot$ and $kT = 1$ keV.

| Source | $F_{X(0.5-2.3 \text{ keV})}$ $(10^{-13} \text{erg cm}^{-2}\text{s}^{-1})$ | $L_{X(0.5-2.3 \text{ keV})}$ $(10^{41} \text{erg s}^{-1})$ |
|--------|---|---|
| HCG 3  | < 1.62 | < 4.54 |
| HCG 16 | < 0.41 | < 0.15 |
| HCG 38 | < 3.20 | < 5.86 |
| HCG 44 | < 0.70 | < 0.03 |
| HCG 57 | < 2.98 | < 5.91 |
| HCG 73 | < 6.02 | < 26.1 |

Table 5: Positions and count rates of the unclassified X-ray sources (see text).

| Source | RA (2000) | Dec (2000) | Count-Rate ($s^{-1}$) |
|---|---|---|---|
| | HCG16 | | |
| XR1 | 02 09 58.4 | -10 03 12.0 | $0.0091 \pm 0.0008$ |
| XR2 | 02 08 56.1 | -10 03 11.6 | $0.0057 \pm 0.0007$ |
| | HCG44 | | |
| XR1 | 10 18 53.9 | 21 51 43.8 | $0.0184 \pm 0.002$ |
| XR2 | 10 18 00.0 | 21 54 22.7 | $0.0197 \pm 0.002$ |
| XR3 | 10 17 44.6 | 21 51 14.2 | $0.0033 \pm 0.001$ |
| XR4 | 10 17 32.7 | 21 51 06.1 | $0.0052 \pm 0.001$ |

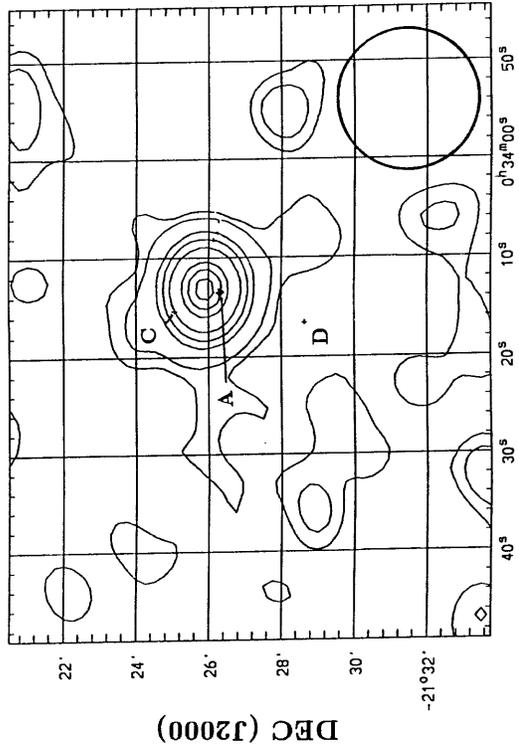

Fig. 3 HCG 12

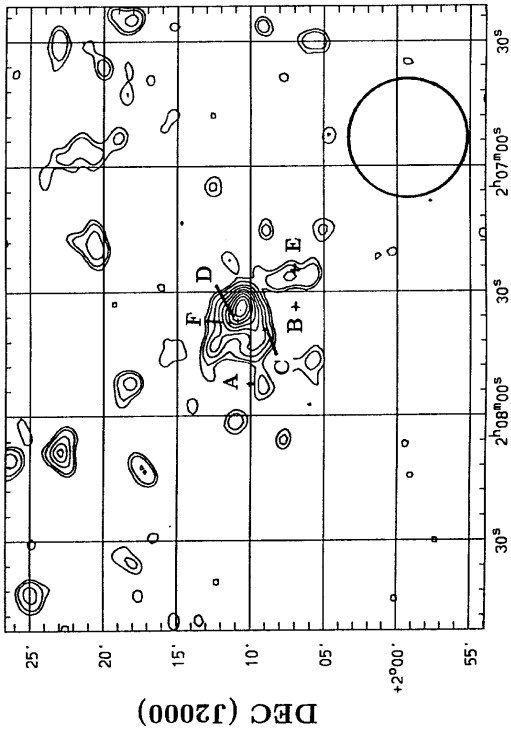

Fig. 5 HCG 16

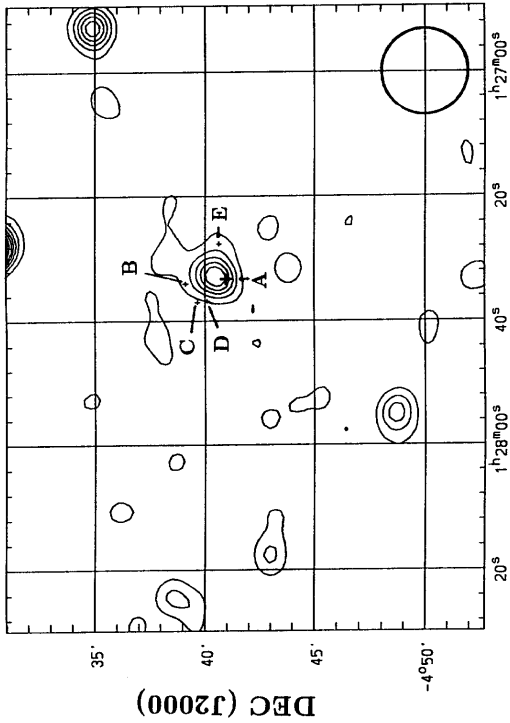

Fig. 1 HCG 4

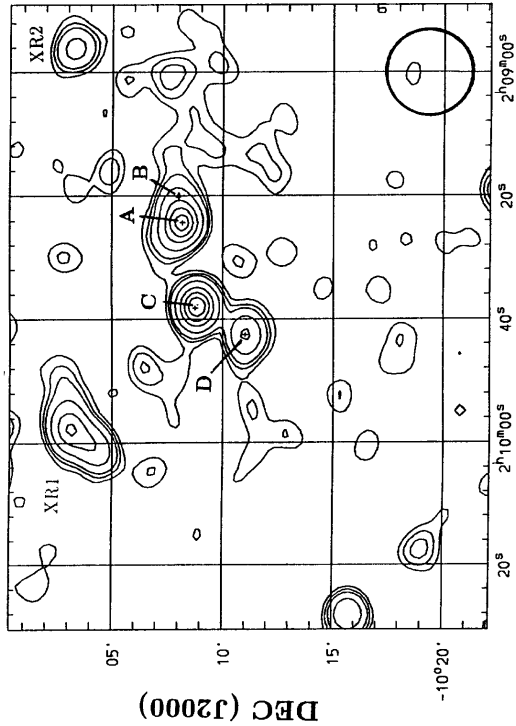

Fig. 4 HCG 15

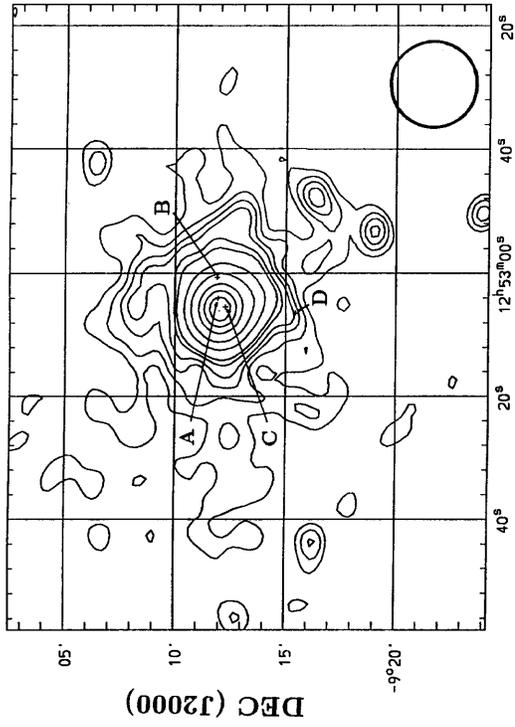

Fig. 7 HCG 62

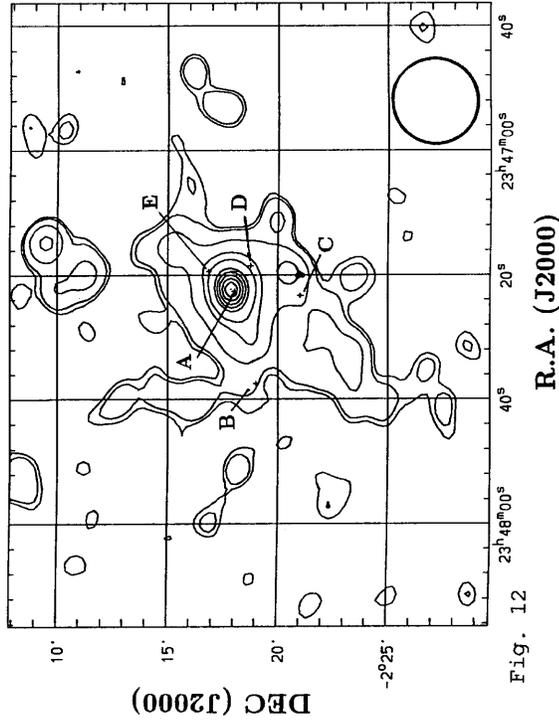

Fig. 12 HCG 97

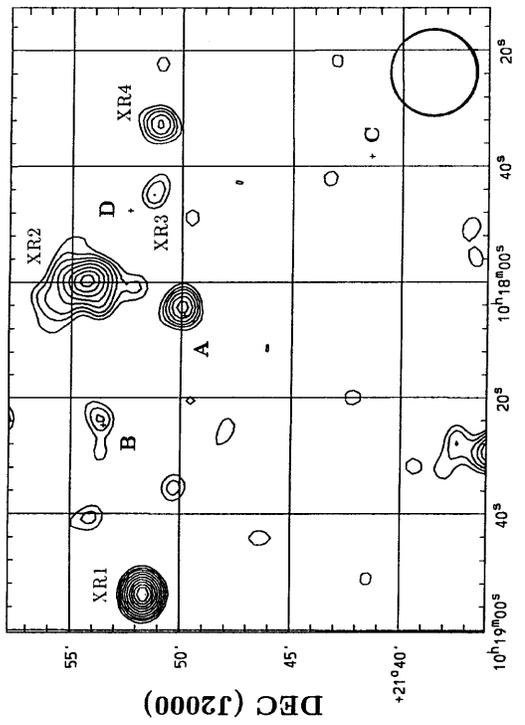

Fig. 6 HCG 44

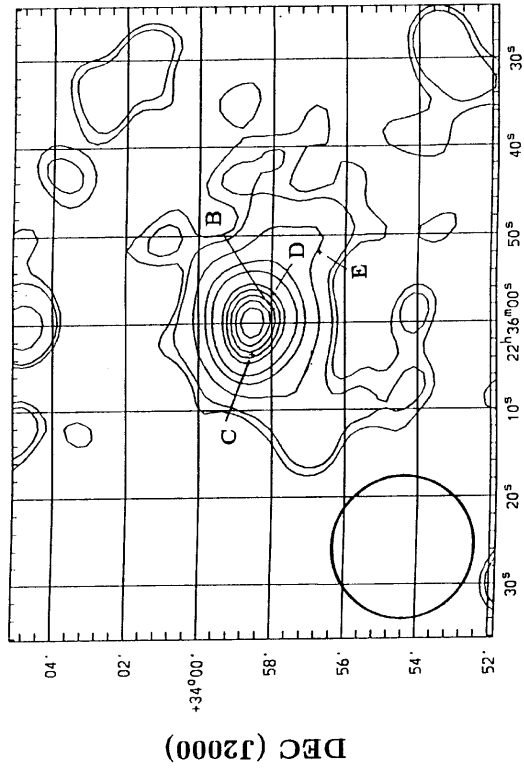

Fig. 10 HCG 92

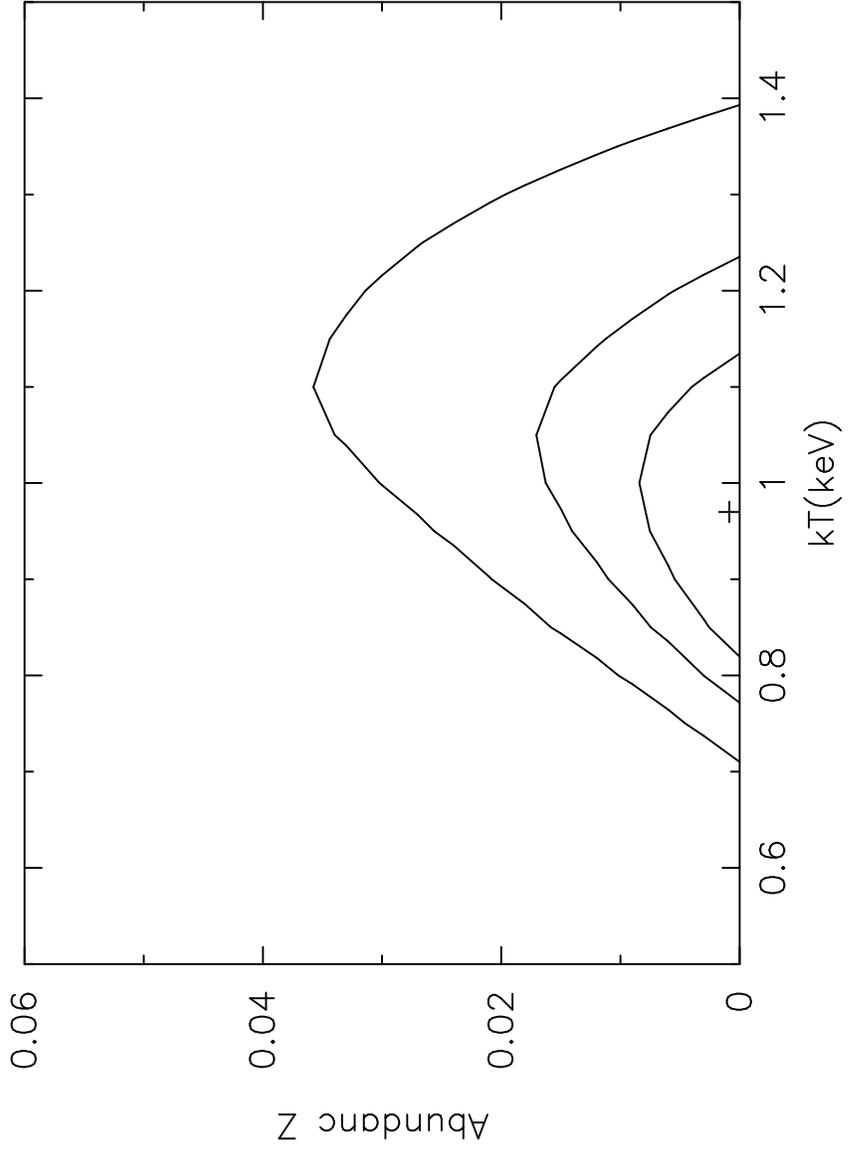

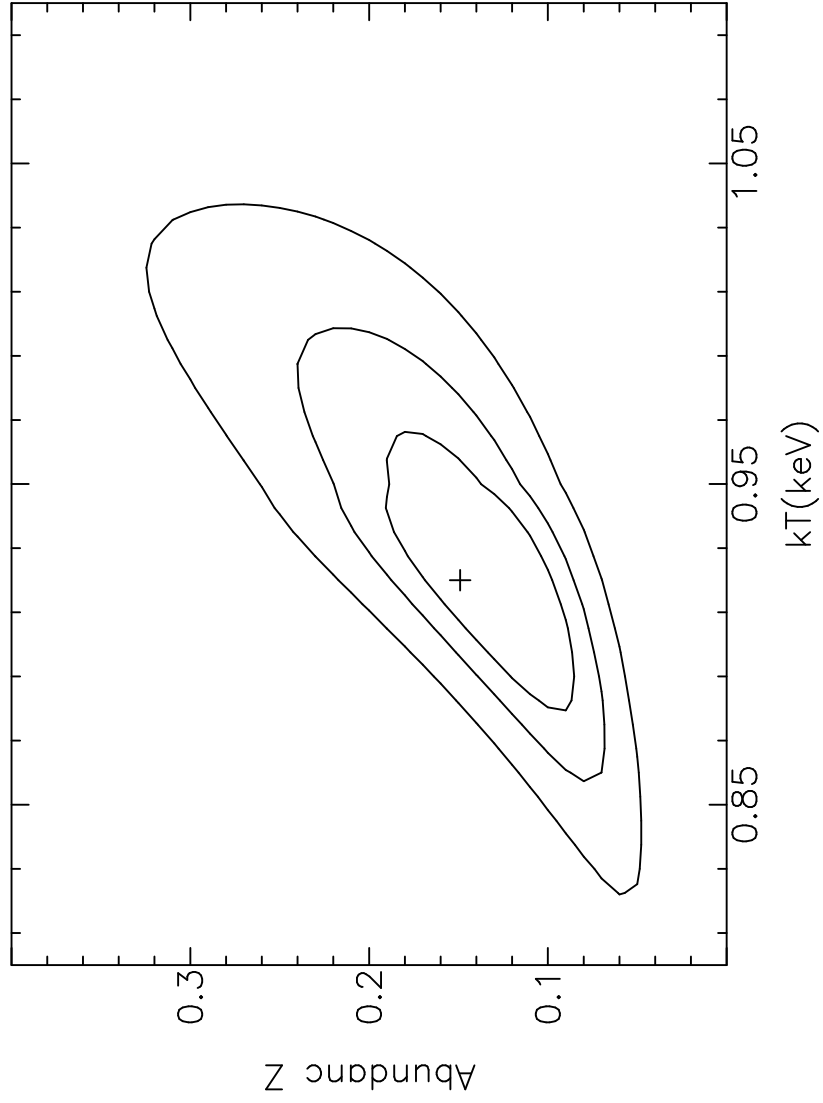

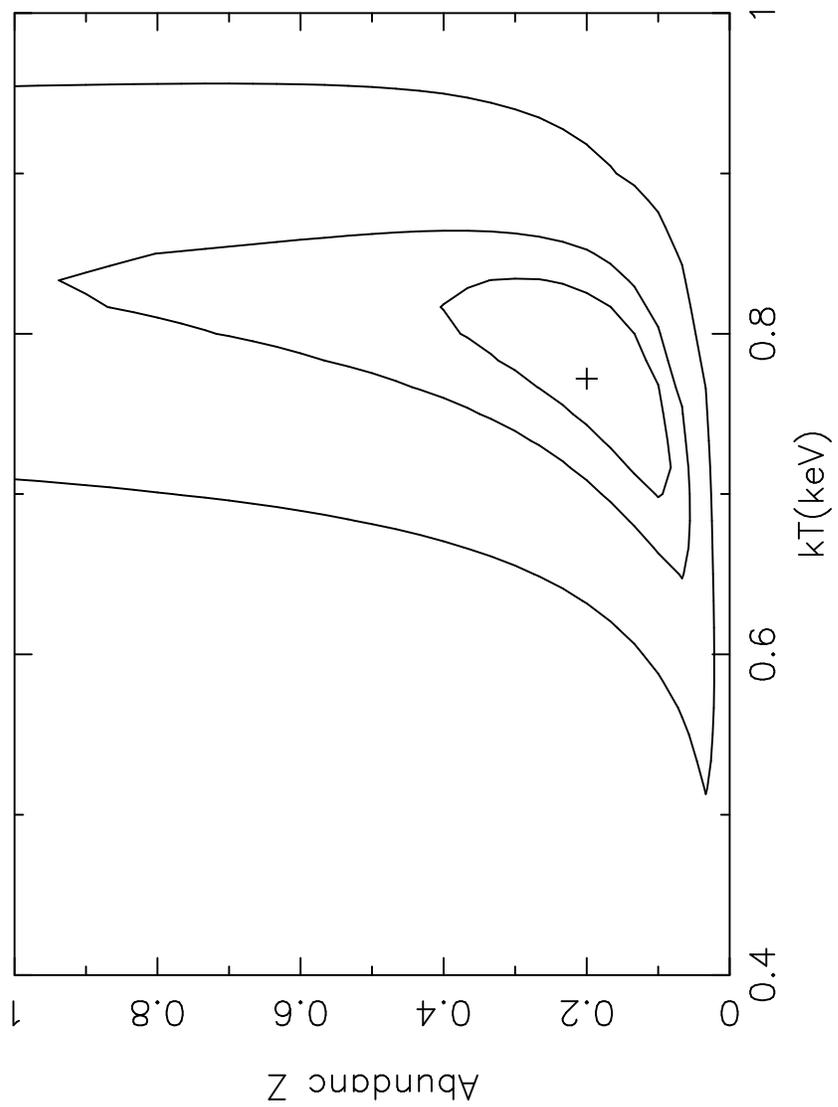

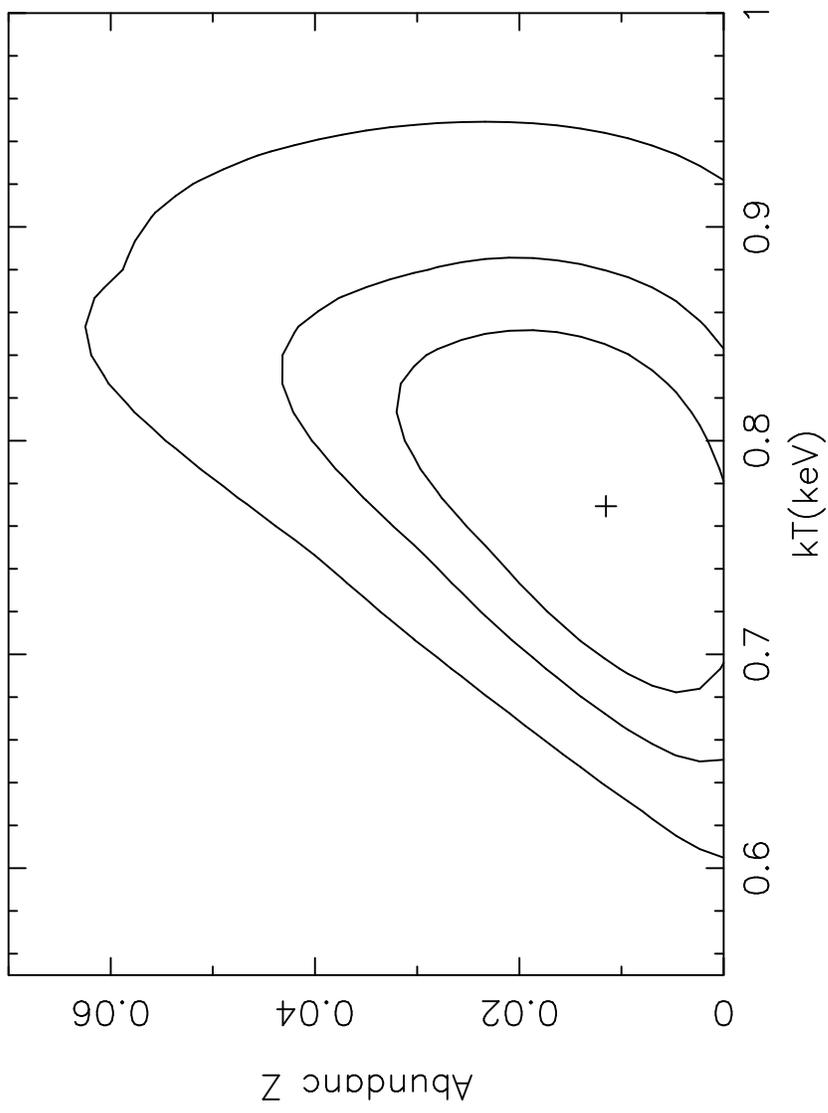

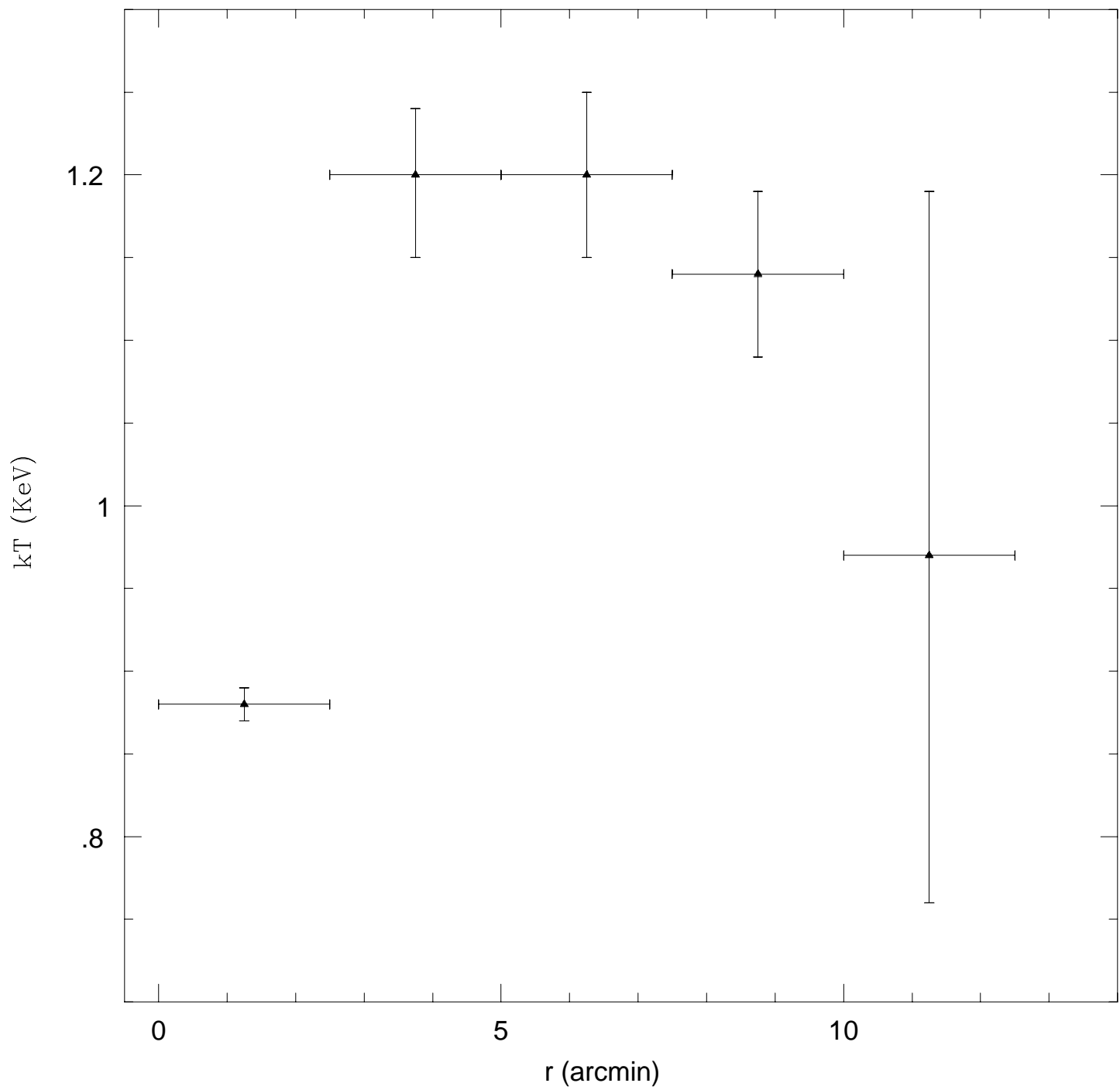

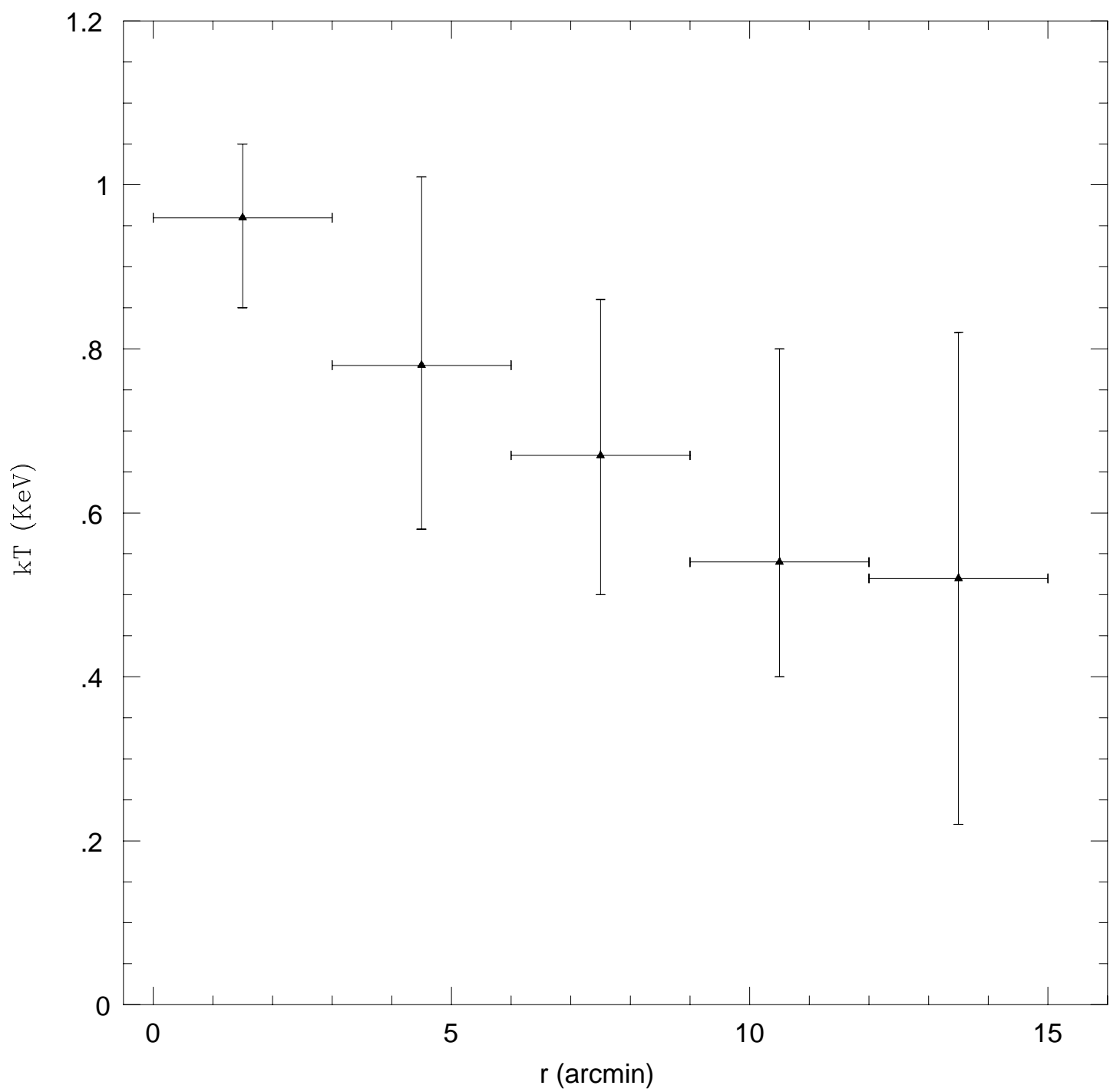